\documentclass[reprint,superscriptaddress,amsmath,amssymb,aps,prl,notitlepage,floatfix,nofootinbib,longbibliography,]{revtex4-1}

\usepackage{tensor}     % manipulate tensors
\usepackage{graphicx}   % include figures
\usepackage[
colorlinks=true,        % color link
citecolor=blue,         % cite color
linkcolor=blue,         % link color
urlcolor=blue           % url color
]{hyperref}             % create hyperlinks
\usepackage{bm}         % \bm{<text>} Bold math symbols
\usepackage{xcolor}     % textcolor
\usepackage{lipsum}
\usepackage{color}      % \textcolor{declared-color}{text}
\usepackage[utf8]{inputenc} % accented characters for .bib file using XeLaTex
\usepackage[section]{placeins} % figures processing
\usepackage{multirow}
\usepackage{orcidlink}
\newcommand{\nc}{\newcommand*} 

\nc{\al}{\alpha}
\nc{\s}{\sigma}
\nc{\kp}{\kappa}
\nc{\dt}{\delta}
\nc{\Dt}{\Delta}
\nc{\Ld}{\Lambda}
\nc{\p}{\partial}
\nc{\Gm}{\Gamma}
\nc{\om}{\omega}
\nc{\Om}{\Omega}
\nc{\rd}{\mathrm{d}}
\def\({\left(}
\def\){\right)}
\def\[{\left[}
\def\]{\right]}
\def\e{\begin{equation}}
\def\q{\end{equation}}
\def\m{\begin{eqnarray}}
\def\n{\end{eqnarray}}
\nc{\Eq}[1]{Eq.~\eqref{#1}}     % equation
\nc{\Fig}[1]{Fig.~\ref{#1}}     % figure
\nc{\Table}[1]{Table~\ref{#1}}  % table
\nc{\Sec}[1]{Sec.~\ref{#1}}     % section
\nc{\Msun}{M_\odot}             % solar mass
\nc{\fpbh}{f_{\mathrm{pbh}}}    % f_pbh
\nc{\fpbhn}{f_{\mathrm{pbh0}}}    % f_pbh
\nc{\mR}{\mathcal{R}} % merger rate density
\nc{\seq}{\sigma_{\mathrm{eq}}}
\nc{\ogw}{\Omega_{\mathrm{GW}}}
\nc{\gpcyr}{\mathrm{Gpc}^{-3}\,\mathrm{yr}^{-1}}
\nc{\lvc}{LIGO/Virgo} % LIGO-VIRGO collaboration
\nc{\SNR}{\mathrm{SNR}} % signal to noise ratio
\nc{\mmin}{{m_{\mathrm{min}}}}
\nc{\mmax}{{m_{\mathrm{max}}}}
\nc{\Mmin}{{M_{\mathrm{min}}}}
\nc{\fmin}{{f_{\mathrm{min}}}}
\nc{\VT}{\mathrm{VT}}
\nc{\rhoGW}{\rho_{\mathrm{GW}}}
\nc{\vth}{\vec{\theta}}
\nc{\vd}{\vec{d}}
\nc{\vla}{\vec{\lambda}}
\nc{\Nobs}{N_{\mathrm{obs}}}
\nc{\av}[1]{\langle #1 \rangle} % average bracket
\nc{\km}{\mathrm{km}}
\nc{\Mpc}{\mathrm{Mpc}}
\nc{\Tobs}{T_{\mathrm{obs}}}
\nc{\Ntemp}{N_{\mathrm{temp}}}
\nc{\fyr}{f_{\mathrm{yr}}}
\nc{\addref}{[\textcolor{red}{add ref}] } % placeholder of references
\nc{\eg}{\textit{e.g.~}}
\nc{\app}{\approx}
\nc{\hf}{\frac{1}{2}}
\nc{\discuss}{\textcolor{red}{Add discussion here!}}
\nc{\red}[1]{\textcolor{red}{#1}}
\nc{\hp}{h_+} % h plus
\nc{\hc}{h_{\times}} % h cross
\nc{\Oh}{\hat{\Omega}}
\nc{\vx}{\vec{x}}
\nc{\mh}{\hat{m}}
\nc{\nh}{\hat{n}}
\nc{\zh}{\hat{z}}
\nc{\ph}{\hat{p}}
\nc{\A}[1]{\mathcal{A}_{#1}}
\nc{\Ogw}[1]{\Omega_{\mathrm{#1}}}
\nc{\bn}[1]{\dt\bm{t}_{\text{#1}}}
\nc{\bC}[1]{\bm{C}_{\text{#1}}}
\nc{\NTOA}{N_{\text{TOA}}}
\nc{\Nmode}{{N_{\text{mode}}}}
\nc{\ARN}{A_{\rm{RN}}}
\nc{\gRN}{\gamma_{\rm{RN}}}
\nc{\bS}{\mathbf{\Sigma}}
\nc{\br}{\mathbf{r}}
\nc{\bN}{\mathbf{R}}
\nc{\Agw}{A_\mathrm{GWB}}
\nc{\UCP}{\mathrm{UCP}}
\nc{\TT}{\mathrm{TT}}
\nc{\ST}{\mathrm{ST}}
\nc{\SL}{\mathrm{SL}}
\nc{\VL}{\mathrm{VL}}

\nc{\BFST}{$107 \pm 7$}
\begin{document}
%%%%%%%%%%%%%%%%%%%%%%%%%%%%%%%%%%%%%%%%%%%%%%%%%%%%%%%%%%%%%%%%%%%%%%%%%%%%%%%%
	
%%%%%%%%%%%%%%%%%%%%%%%%%%%%%%%%%%%% title %%%%%%%%%%%%%%%%%%%%%%%%%%%%%%%%%%%%%
\title{Search for Non-Tensorial Gravitational-Wave Backgrounds in the NANOGrav\\ 15-Year Data Set}	
	
%%%%%%%%%%%%%%%%%%%%%%%%%%%%%%%%%%%% author %%%%%%%%%%%%%%%%%%%%%%%%%%%%%%%%%%%%
\author{Zu-Cheng Chen\orcidlink{0000-0001-7016-9934}}
\email{zuchengchen@hunnu.edu.cn}
\affiliation{Department of Physics and Synergetic Innovation Center for Quantum Effects and Applications, Hunan Normal University, Changsha, Hunan 410081, China}
\affiliation{Institute of Interdisciplinary Studies, Hunan Normal University, Changsha, Hunan 410081, China}
\affiliation{Department of Astronomy, Beijing Normal University, Beijing 100875, China}
\affiliation{Advanced Institute of Natural Sciences, Beijing Normal University, Zhuhai 519087, China}

%%%%%%%%%%%%%%%%%%%%%%%%%%%%%%%%%%%% author %%%%%%%%%%%%%%%%%%%%%%%%%%%%%%%%%%%%
\author{Yu-Mei Wu\orcidlink{0000-0002-9247-5155}}
\email{Corresponding author: ymwu@ucas.ac.cn} 
\affiliation{School of Fundamental Physics and Mathematical Sciences, Hangzhou Institute for Advanced Study, UCAS, Hangzhou 310024, China}
\affiliation{School of Physical Sciences, University of Chinese Academy of Sciences, No. 19A Yuquan Road, Beijing 100049, China}
% \affiliation{CAS Key Laboratory of Theoretical Physics, Institute of Theoretical Physics, Chinese Academy of Sciences, Beijing 100190, China}

%%%%%%%%%%%%%%%%%%%%%%%%%%%%%%%%%%%% author %%%%%%%%%%%%%%%%%%%%%%%%%%%%%%%%%%%%
\author{Yan-Chen Bi\orcidlink{0000-0002-9346-8715}}
\email{Corresponding author: biyanchen@itp.ac.cn}
\affiliation{School of Physical Sciences, 
    University of Chinese Academy of Sciences, 
    No. 19A Yuquan Road, Beijing 100049, China}
\affiliation{CAS Key Laboratory of Theoretical Physics, 
    Institute of Theoretical Physics, Chinese Academy of Sciences, Beijing 100190, China}
    
%%%%%%%%%%%%%%%%%%%%%%%%%%%%%%%%%%%% author %%%%%%%%%%%%%%%%%%%%%%%%%%%%%%%%%%%%
\author{Qing-Guo Huang\orcidlink{0000-0003-1584-345X}}
\email{Corresponding author: huangqg@itp.ac.cn}
\affiliation{School of Fundamental Physics and Mathematical Sciences, Hangzhou Institute for Advanced Study, UCAS, Hangzhou 310024, China}
\affiliation{School of Physical Sciences, 
    University of Chinese Academy of Sciences, 
    No. 19A Yuquan Road, Beijing 100049, China}
\affiliation{CAS Key Laboratory of Theoretical Physics, 
    Institute of Theoretical Physics, Chinese Academy of Sciences,
    Beijing 100190, China}

%%%%%%%%%%%%%%%%%%%%%%%%%%%%%%%%% abstract %%%%%%%%%%%%%%%%%%%%%%%%%%%%%%%%%%%%%
\begin{abstract}
The recent detection of a stochastic signal in the NANOGrav 15-year data set has aroused great interest in uncovering its origin. 
However, the evidence for the Hellings-Downs correlations, a key signature of the gravitational-wave background (GWB) predicted by general relativity, remains inconclusive.
In this letter, we search for an isotropic non-tensorial GWB, allowed by general metric theories of gravity, in the NANOGrav 15-year data set. 
Our analysis reveals a Bayes factor of approximately 2.5, comparing the quadrupolar (tensor transverse, TT) correlations to the scalar transverse (ST) correlations, suggesting that the ST correlations provide a comparable explanation for the observed stochastic signal in the NANOGrav data. We obtain the median and the $90\%$ equal-tail amplitudes as $\mathcal{A}_\mathrm{ST} = 7.8^{+5.1}_{-3.5} \times 10^{-15}$ at the frequency of 1/year. 
Furthermore, we find that the vector longitudinal (VL) and scalar longitudinal (SL) correlations are weakly and strongly disfavoured by data, respectively, yielding upper limits on the amplitudes: $\mathcal{A}_\mathrm{VL}^{95\%} \lesssim 1.7 \times 10^{-15}$ and $\mathcal{A}_\mathrm{SL}^{95\%} \lesssim 7.4 \times 10^{-17}$. Lastly, we fit the NANOGrav data with the general transverse (GT) correlations parameterized by a free parameter $\alpha$. Our analysis yields $\alpha=1.74^{+1.18}_{-1.41}$, thus excluding both the TT ($\alpha=3$) and ST ($\alpha=0$) models at the $90\%$ confidence level.
\end{abstract}

\maketitle

%%%%%%%%%%%%%%%%%%%%%%%%%%%%%%%%%%%%%%%%%%%%%%%%%%%%%%%%%%%%%%%%%%%%%%%%%%%%%%%%
\textbf{Introduction.}
A pulsar timing array (PTA) is dedicated to the detection of gravitational waves (GWs) with frequencies in the nanohertz range by regularly monitoring the spatially correlated fluctuations caused by GWs on the time of arrivals (TOAs) of radio pulses emitted by an array of pulsars~\cite{1978SvA....22...36S,Detweiler:1979wn,1990ApJ...361..300F}.
There are three major PTA projects: the European PTA (EPTA)~\cite{Kramer:2013kea}, the North American Nanoherz Observatory for GWs (NANOGrav)~\cite{McLaughlin:2013ira}, and the Parkes PTA (PPTA)~\cite{Manchester:2012za}. Over the course of more than a decade, these projects have been monitoring the TOAs from dozens of millisecond pulsars with an observation cadence ranging from weekly to monthly. These PTAs along with the Indian PTA (InPTA)~\cite{2018JApA...39...51J} constitute the International PTA (IPTA)~\cite{Hobbs:2009yy,Manchester:2013ndt}. Meanwhile, the Chinese PTA (CPTA)~\cite{2016ASPC..502...19L} and the MeerKAT PTA (MPTA)~\cite{Miles:2022lkg} are relatively new PTA collaborations that use the sensitive new telescopes, FAST and MeerKAT.

Recently, NANOGrav~\cite{NANOGrav:2023hde,NANOGrav:2023gor}, EPTA+InPTA~\cite{Antoniadis:2023lym,Antoniadis:2023ott}, PPTA~\cite{Zic:2023gta,Reardon:2023gzh}, and CPTA~\cite{Xu:2023wog} have independently announced compelling evidence for a stochastic signal in their latest data sets. These data sets demonstrate varying levels of significance in supporting the presence of Hellings-Downs (HD)~\cite{Hellings:1983fr} spatial correlations as predicted by general relativity. 
While the PTA window covers a broad range of possible sources~\cite{Li:2019vlb,Chen:2019xse,Vagnozzi:2020gtf,Benetti:2021uea,Chen:2022azo,Ashoorioon:2022raz,PPTA:2022eul,Wu:2023pbt,IPTA:2023ero,Wu:2023dnp,Dandoy:2023jot,Madge:2023cak,Bi:2023ewq,Wu:2023rib}, the exact origin of the observed signal remains under active investigation, whether from astrophysical phenomena or cosmological processes~\cite{NANOGrav:2023hvm,Antoniadis:2023xlr,King:2023cgv,Niu:2023bsr,Ben-Dayan:2023lwd,Vagnozzi:2023lwo,Fu:2023aab,InternationalPulsarTimingArray:2023mzf,Basilakos:2023jvp}.
A variety of sources can potentially explain the PTA signal~\cite{Wu:2023hsa,Ellis:2023oxs,Figueroa:2023zhu}, including the GW background (GWB) generated by supermassive black hole binaries~\cite{NANOGrav:2023hfp,Ellis:2023dgf,Bi:2023tib}, domain walls~\cite{Kitajima:2023cek,Babichev:2023pbf}, cosmic strings~\cite{Kitajima:2023vre,Ellis:2023tsl,Ahmed:2023pjl}, phase transitions~\cite{Addazi:2023jvg,Athron:2023mer,Xiao:2023dbb,Abe:2023yrw,Gouttenoire:2023bqy}, and scalar-induced GWs~\cite{Franciolini:2023pbf,Jin:2023wri,Liu:2023pau,Yi:2023npi,Harigaya:2023pmw,Liu:2023ymk,Liu:2023hpw} accompanying the formation of primordial black holes~\cite{Chen:2018czv,Bousder:2023ida,Gouttenoire:2023nzr}.

Identifying a GWB as predicted by general relativity hinges on the observation of its quintessential quadrupolar characteristics, specifically the HD spatial correlations within PTA data.
To achieve this goal, it is crucial to conduct a consistency test to confirm that the signal exhibits clearly quadrupolar characteristics~\cite{Allen:2023kib}, thereby ruling out other reasonable explanations such as the monopolar or dipolar correlations. 
While the NANOGrav 15-year data set strongly disfavors individual monopole and dipole signals and exhibits a slight disfavoring tendency toward HD+monopole and HD+dipole models~\cite{NANOGrav:2023gor}, it is important to note that this does not exclude the possibility of alternative GW polarization modes allowed in general metric theories of gravity. In fact, a most general metric gravity theory can have two scalar modes and two vector modes in addition to the two tensor modes, each with distinct correlation patterns~\cite{2008ApJ...685.1304L,Chamberlin:2011ev,Gair:2015hra,Boitier:2020xfx,Bernardo:2022vlj,Bernardo:2022rif}.
It is worth noting that earlier studies~\cite{Chen:2021wdo,NANOGrav:2021ini,Wu:2021kmd,Chen:2021ncc} have tentatively reported evidence for scalar transverse (ST) correlations.
To determine whether the observed PTA signal indeed originates from a GWB as predicted by general relativity, it is imperative to fit the data with all plausible correlation patterns. In this letter, we perform the Bayesian search for the stochastic GWB signal, modelled by a power-law spectrum with a varying power-law index. Our analysis considers all six polarization modes in the NANOGrav 15-year data set. 

%%%%%%%%%%%%%%%%%%%%%%%%%%%%%%%%%%%%%%%%%%%%%%%%%%%%%%%%%%%%%%%%%%%%%%%%%%%%%%%%
\textbf{Detecting non-tensorial GWBs with PTAs.}
Pulsar timing experiments take advantage of the regular arrival rates of radio pulses emitted by extremely stable millisecond pulsars. GWs can perturb the geodesics of these radio waves, leading to fluctuations in the TOAs of radio pulses~\cite{1978SvA....22...36S,Detweiler:1979wn}. The presence of a GW will result in unexplained residuals in the TOAs, even after compensating for a deterministic timing model that accounts for the pulsar spin behaviour and the geometric effects caused by the motion of the pulsar and the Earth~\cite{1978SvA....22...36S,Detweiler:1979wn}. 
Through regularly monitoring TOAs of pulsars from an array of the highly stable millisecond pulsars~\cite{1990ApJ...361..300F}, and analyzing the expected cross-correlations among pulsars in a PTA, it becomes possible to extract the GW signal from other systematic effects, such as the clock errors.

\begin{figure}[tbp!]
    \centering
    \includegraphics[width=\linewidth]{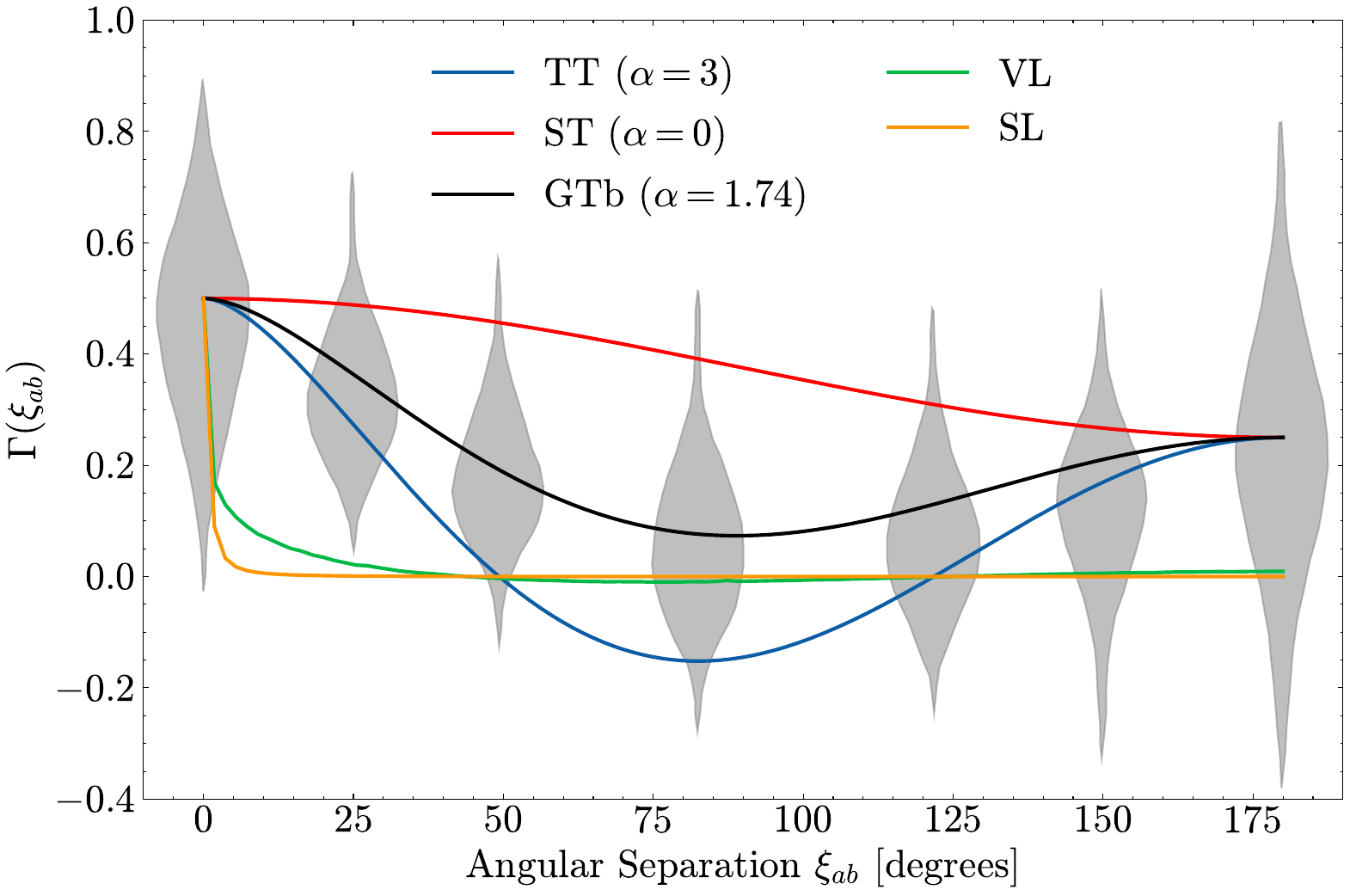}
    \caption{\label{orfs}Various ORFs examined in this work.  Here, the GTb model refers to the GT model with $\alpha=1.74$. The Bayesian reconstruction of normalized inter-pulsar correlations, as adapted from NANOGrav's analysis~\cite{NANOGrav:2023gor}, is depicted by the gray violins. It is important to emphasize that the longitudinal modes, VL and SL, depend on both pulsar distance and GW frequency. The illustration here specifically presents the scenario where $fL_a = fL_b = 100$ for the VL and SL modes.}
\end{figure}

The cross-power spectral density of timing residuals induced by a GWB at the frequency $f$ for two pulsars, $a$ and $b$, can be expressed as~\cite{2008ApJ...685.1304L,Chamberlin:2011ev,Gair:2015hra}
\e\label{Sab1}
S_{ab}(f) = \sum_P \frac{h_{c,P}^2}{12 \pi^2 f^3} \Gm^P_{ab}(f).
\q 
Here, $h_c^P(f)$ represents the characteristic strain, and the summation encompasses all six possible GW polarizations that can be inherent in a general metric gravity theory, specifically denoted as $P = +, \times, x, y, l, b$.
The symbols ``$+$" and ``$\times$" refer to the two spin-2 transverse traceless polarization modes; ``$x$" and ``$y$" correspond to the two spin-1 shear modes; ``$l$" designates the spin-0 longitudinal mode; and ``$b$" identifies the spin-0 breathing mode. 
The overlap reduction function (ORF) $\Gm^P_{ab}$ for a pair of pulsars is given by~\cite{2008ApJ...685.1304L,Chamberlin:2011ev}
\begin{align}
\Gm^P_{ab}(f) =& \frac{3}{8\pi} \int d\Oh \(e^{2\pi i f L_a(1+\Oh\cdot\ph_a)}-1\)\nonumber\\
&\times\(e^{2\pi i f L_b(1+\Oh\cdot\ph_b)}-1\) F^P_a(\Oh) F^P_b(\Oh),
\end{align}
where $\ph$ is the direction of the pulsar with respect to the Earth, $L_a$ and $L_b$ are the distance from the Earth to the pulsar $a$ and $b$ respectively, and $\Oh$ is the propagating direction of the GW. Additionally, antenna patterns $F^P(\Oh)$ are expressed as
\e 
F^P(\Oh) = e^P_{ij}(\Oh) \frac{\ph^i\ph^j}{2(1+\Oh\cdot\ph)},
\q
where $e^P_{ij}$ stands for the polarization tensor corresponding to polarization mode $P$~\cite{2008ApJ...685.1304L,Chamberlin:2011ev}. As per the conventions established in~\cite{Cornish:2017oic}, we define
\begin{align}
\Gm^{\TT}_{ab}(f) &= \Gm^{+}_{ab}(f) + \Gm^{\times}_{ab}(f), \label{gammaTT}\\
\Gm^{\ST}_{ab}(f) &= \Gm^{b}_{ab}(f), \label{gammaST}\\
\Gm^{\VL}_{ab}(f) &= \Gm^{x}_{ab}(f) + \Gm^{y}_{ab}(f), \label{gammaVL}\\
\Gm^{\SL}_{ab}(f) &= \Gm^{l}_{ab}(f). \label{gammaSL}
\end{align}
For the tensor transverse (TT) and $\ST$ polarization modes, the ORFs exhibit a notable property of being nearly independent of both distance and frequency, which can be analytically computed by~\cite{Hellings:1983fr,2008ApJ...685.1304L}
\m\label{TTST}
\Gm^{\TT}_{ab}(f) &=& \hf(1+\dt_{ab})+ \frac{3}{2} k_{ab} \(\ln k_{ab}-\frac{1}{6}\), \\
\Gm^{\ST}_{ab}(f) &=& \frac{1}{8}\(3 + 4\dt_{ab} + \cos\zeta_{ab}\),
\n 
where $\dt_{ab}$ represents the Kronecker delta symbol, $\zeta_{ab}$ denotes the angular separation between pulsars $a$ and $b$, and $k_{ab} \equiv (1-\cos\zeta_{ab})/2$. Note that $\Gm^{\TT}_{ab}$ is commonly referred to as HD correlations, which are closely associated with the quadrupolar nature of GW signals.
In contrast, analytical expressions for the vector longitudinal (VL) and scalar longitudinal (SL) polarization modes are not readily available. Therefore, we rely on numerical methods to compute these functions. In this work, we adopt the pulsar distance information collected in Table~2 of~\cite{NANOGrav:2023pdq} to estimate the ORFs.
It's worth noting that the ORFs for $\mathrm{TT}$ and $\mathrm{ST}$ polarization modes only differ by the presence or absence of the term $\kappa_{ab}\ln \kappa_{ab}$. To generalize these ORFs, we adopt a parameterized form, following a similar approach as described in~\cite{Chen:2021ncc},
\e\label{alpha}
\Gamma_{ab}=\frac{1}{8}(3+4\delta_{ab}+\cos\xi_{ab})+\frac{\alpha}{2}\kappa_{ab}\ln \kappa_{ab}.
\q
Here, the parameter $\alpha$ allows us to seamlessly transition between the $\mathrm{TT}$ mode (when $\alpha=3$) and the $\mathrm{ST}$ mode (when $\alpha=0$).
For later convenience, we refer to this parameterization as the ``general transverse" (GT). See \Fig{orfs} for a visualization of various ORFs considered in this work.

%%%%%%%%%%%%%%%%%%%%%%%%%%%%%%%%%%%%%%%%%%%%%%%%%%%%%%%%%%%%%%%%%%%%%%%%%%%%%%%%

Since current PTA data is not able to distinguish between various spectral shapes of the GWB energy density~\cite{NANOGrav:2023hvm,Wu:2023hsa}, we employ a power-law energy density spectrum in our analysis. This leads to the following expression:
\e\label{Sab2}
    S_{ab}(f) = \sum_{I={\TT, \ST, \VL, \SL}} \Gm^I_{ab} \frac{\A{I}^2}{12\pi^2}   \(\frac{f}{\fyr}\)^{-\gamma_I} \fyr^{-3},
\q
where $\A{I}$ is the GWB amplitude of polarization mode $I$, $\fyr = 1/\mathrm{year}$, and $\gamma_I$ corresponds to the spectral index for the polarization mode $I$ that we treat as a free parameter.
The dimensionless GW energy density parameter per logarithm frequency for the polarization mode $I$ is related to $\A{I}$ by,~\cite{Thrane:2013oya}, 
\e 
    \ogw^{I}(f) = \frac{2\pi^2}{3 H_0^2} f^2 h_{c,I}^2 = \frac{2\pi^2\fyr^2}{3 H_0^2} \A{I}^2 \(\frac{f}{\fyr}\)^{5-\gamma_I},
\q 
where $H_0 = 67.4\, \km\, \mathrm{s}^{-1} \Mpc^{-1}$~\cite{Planck:2018vyg} is the Hubble constant.
 
%%%%%%%%%%%%%%%%%%%%%%%%%%%%%%%%%%%%%%%%%%%%%%%%%%%%%%%%%%%%%%%%%%%%%%%%%%%%%%%%
\begin{table*}[tbp!]
    %\begin{center}
    % \scriptsize
    \begin{tabular}{llll}
        \hline\hline
        \textbf{Parameter}\hspace{2mm} & \textbf{Description} & \textbf{Prior} & \textbf{Comments} \\
        \hline
        \multicolumn{4}{c}{White Noise} \\[1pt]
        $E_{k}$ & EFAC per backend/receiver system & Uniform $[0, 10]$ & single-pulsar analysis only \\
        $Q_{k}$[s] & EQUAD per backend/receiver system & log-Uniform $[-8.5, -5]$ & single-pulsar analysis only \\
        $J_{k}$[s] & ECORR per backend/receiver system\hspace{2mm} & log-Uniform $[-8.5, -5]$ & single-pulsar analysis only \\
        \hline
        % \vspace{0.5em}
        \multicolumn{4}{c}{Red Noise} \\[1pt]
        $\ARN$ & red-noise power-law amplitude & log-Uniform $[-20, -11]$\hspace{2mm} & one parameter per pulsar  \\
        $\gRN$ & red-noise power-law spectral index & Uniform $[0, 7]$ & one parameter per pulsar \\
        \hline
        \multicolumn{4}{c}{GWB Process} \\[1pt]
        $A_I$ & GWB amplitude of polarization $I$ & log-Uniform $[-18, -11]$ & one parameter for PTA \\
        $\gamma_{I}$ & power-law index of polarization $I$ & Uniform $[0, 7]$ & one parameter for PTA \\
        \hline
    \end{tabular}
    \caption{\label{tab:priors}Parameters and their prior distributions used in the analyses.}
\end{table*}
%%%%%%%%%%%%%%%%%%%%%%%%%%%%%%%%%%%%%%%%%%%%%%%%%%%%%%%%%%%%%%%%%%%%%%%%%%%%%%%%

%%%%%%%%%%%%%%%%%%%%%%%%%%%%%%%%%%%%%%%%%%%%%%%%%%%%%%%%%%%%%%%%%%%%%%%%%%%%%%%%
\textbf{Data analysis.}
The NANOGrav 15-year data set~\cite{NANOGrav:2023hde} comprises data from $68$ pulsars. 
Following~\cite{NANOGrav:2023gor}, we use $67$ pulsars with timing baselines that exceed three years. 
The timing residuals for each pulsar, obtained by subtracting the timing model from the TOAs, can be expressed as~\cite{Arzoumanian:2015liz}
\e\label{dt}
    \dt\bm{t} = M \bm{\epsilon} + F \bm{a} + \bm{n}.
\q
The term $M \bm{\epsilon}$ serves to accommodate inaccuracies that can arise during the subtraction of the timing model, where $M$ represents the design matrix of the timing model, and $\bm{\epsilon}$ is a vector that denotes minor deviations of the timing model parameters. 
The term $F \bm{a}$ encompasses all low-frequency signals, including both the red noise that is intrinsic to each pulsar and the common red noise signal shared among all pulsars, such as a GWB. Here, $F$ corresponds to the Fourier design matrix, which features components of alternating sine and cosine functions. Given the timespan $T$, $\bm{a}$ is a vector that signifies the amplitude of the Fourier basis functions, and these functions are associated with specific frequencies of $\{1/T, 2/T, \cdots, \Nmode/T\}$.
Similar to NANOGrav~\cite{NANOGrav:2023gor}, we employ $30$ frequency components to account for the intrinsic red noise specific to each pulsar, and these are characterized by a power-law spectrum.
Additionally, we utilize $14$ frequency components for the GWB signal. 
The final term $\bm{n}$ is responsible for modelling the timing residuals stemming from white noise, including a scale parameter on the TOA uncertainties (EFAC), an added variance (EQUAD), and a per-epoch variance (ECORR) for each backend/receiver system~\cite{Arzoumanian:2015liz}.

Similar to NANOGrav~\cite{NANOGrav:2023gor}, we adopt the JPL solar system ephemeris (SSE) DE440~\cite{Park_2021} as the fiducial SSE. 
Our Bayesian parameter inference follows a procedure closely aligned with the one outlined in~\cite{Arzoumanian:2018saf,Arzoumanian:2020vkk}. The model parameters and their associated prior distributions are summarized in~\Table{tab:priors}. 
In our analyses, we keep the white noise parameters fixed at their maximum likelihood values to reduce the computational costs.
We use \texttt{enterprise}~\cite{enterprise} and \texttt{enterprise\_extension}~\cite{enterprise_extensions} software packages for the calculation of likelihood and Bayes factors. The Bayes factors are calculated using the \textit{product-space} method~\cite{10.2307/2346151,10.2307/1391010,Hee:2015eba,Taylor:2020zpk}. For Markov chain Monte Carlo sampling, we utilize the \texttt{PTMCMCSampler}~\cite{justin_ellis_2017_1037579} package. 
To expedite the burn-in process for the chains, we employ samples drawn from empirical distributions to handle the red noise parameters of the pulsars. These distributions are constructed based on posteriors obtained from an initial Bayesian analysis that exclusively incorporates the pulsars' red noise, excluding any common red noise processes, as carried out in~\cite{Aggarwal:2018mgp,Arzoumanian:2020vkk}.

\begin{figure}[tbp!]
    \centering
    \includegraphics[width=\linewidth]{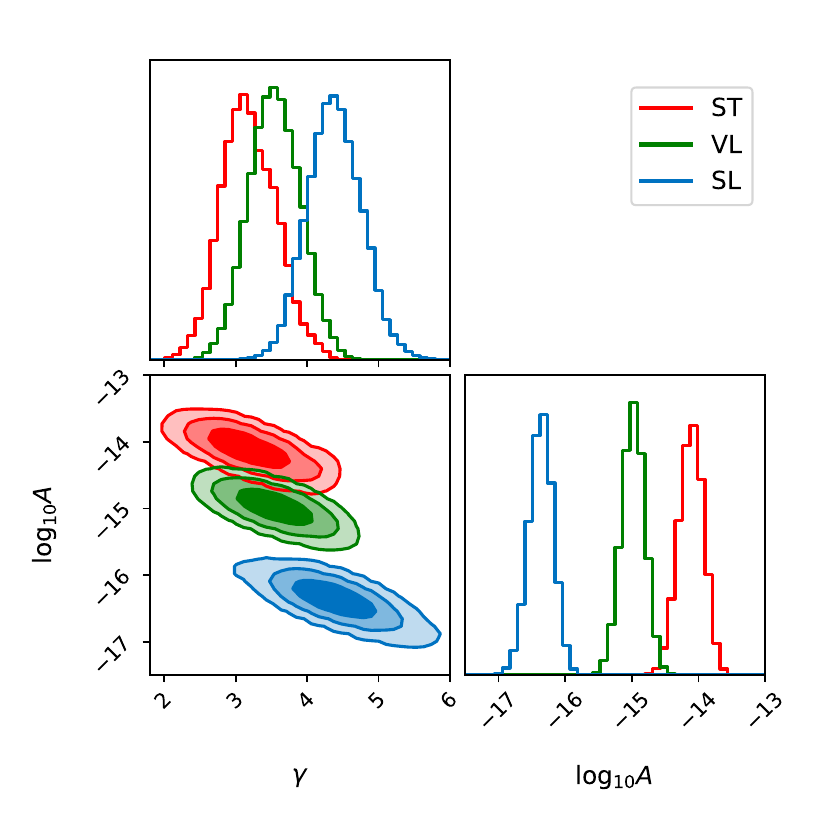}
    \caption{\label{post_ST_VL_SL} Bayesian posteriors for the amplitude, $A$, and power-law index parameter, $\gamma$, obtained in the ST, VL, and SL models. We show the $1\sigma$, $2\sigma$, and $3\sigma$ contours in the two-dimensional plot.}
\end{figure}

\textbf{Results and Discussion.}
\Table{bayes} summarizes the Bayes factors for different models compared to the TT model that incorporates the full HD spatial correlations. The Bayes factor of the ST model relative to the TT model is $0.40\pm 0.03$, indicating that there is no statistically significant evidence supporting or refuting the ST correlations over the HD correlations in the NANOGrav 15-year data set. We obtain the median and the $90\%$ equal-tail amplitudes as $\A{\ST} = 7.8^{+5.1}_{-3.5} \times 10^{-15}$ at the reference frequency of $f_{\rm{yr}}$. The posterior distributions for the amplitude $A_{\ST}$ and the power-law index $\gamma_{\ST}$ are illustrated in \Fig{post_ST_VL_SL}.

%%%%%%%%%%%%%%%%%%%%%%%%%%%%%%%%%%%%%%%%%%%%%%%%%%%%%%%%%%%%%%%%%%%%%%%%%%%%%%%%
\begin{table*}[tbp!]
\begin{center}
\begin{tabular}{c|ccccccccccc}
\hline\hline
Model & ST  & VL & SL  & GTb & TT + ST & TT+VL & TT+SL & TT+Mono & GTb+Mono & TT+Mono+Dipole & GTb+Mono+Dipole\\
\hline
BF & $0.40 (3)$& $0.12 (2)$ & $0.002 (1)$ & $3.9 (3)$ & $0.943 (5)$  & $0.489(6)$ & $0.266(4)$ & $0.548(6)$ & $2.3(6)$ & $0.255(4)$ & $0.26(6)$\\
\hline
\end{tabular}	
\end{center}  
\caption{\label{bayes}The Bayes factors for various models compared to the TT model that considers the full HD spatial correlations derived from the product-space method. Here, the GTb model stands for the GT model with $\alpha=1.74$. For all models, we use the power-law spectrum for the GWB with a varied power-law index parameter. The digit in parentheses gives the uncertainty on the last quoted digit.}
\end{table*}
%%%%%%%%%%%%%%%%%%%%%%%%%%%%%%%%%%%%%%%%%%%%%%%%%%%%%%%%%%%%%%%%%%%%%%%%%%%%%%%%

%%%%%%%%%%%%%%%%%%%%%%%%%%%%%%%%%%%%%%%%%%%%%%%%%%%%%%%%%%%%%%%%%%%%%%%%%%%%%%%%
\begin{table}[tbp!]
\begin{center}
\begin{tabular}{ccccc}
\hline\hline
$\mathrm{BF}^{\mathrm{TT+ST}}_{\mathrm{GTb}}$ & $\mathrm{BF}^{\mathrm{TT+ST}}_{\mathrm{TT+mono}}$ & $\mathrm{BF}^{\mathrm{GTb+mono}}_{\mathrm{TT+mono}}$  & $\mathrm{BF}^{\mathrm{GTb+mono+dipole}}_{\mathrm{TT+mono+dipole}}$\\
\hline
$0.24$ & $1.72$& $4.2$ & $1.02$\\
\hline
\end{tabular}	
\end{center}  
\caption{\label{bayes2}The Bayes factors between various models by a conversion from \Table{bayes}, \textit{e.g.} $\mathrm{BF}^{\mathrm{TT+ST}}_{\mathrm{GTb}} = \mathrm{BF}^{\mathrm{TT+ST}}_{\mathrm{TT}}/ \mathrm{BF}^{\mathrm{GTb}}_{\mathrm{TT}}$.}
\end{table}
%%%%%%%%%%%%%%%%%%%%%%%%%%%%%%%%%%%%%%%%%%%%%%%%%%%%%%%%%%%%%%%%%%%%%%%%%%%%%%%%

The Bayes factor of the VL model versus the TT model is $0.12\pm 0.02$, implying the VL correlations are slightly disfavoured in comparison to the HD correlations. Furthermore, the Bayes factor of the SL model compared to the TT model is $0.002\pm 0.001$, strongly disfavoring the SL correlations compared to the HD correlations. Consequently, we establish upper limits for the amplitudes as $\A{\VL} \lesssim 1.7 \times 10^{-15}$ and $\A{\SL} \lesssim 7.4 \times 10^{-17}$. Note that the constraint on the amplitude for the SL model is around two orders of magnitude tighter than that from other polarizations, mainly due to the strong auto-correlations inherent to the SL mode. The posteriors for the VL and SL models are also shown in \Fig{post_ST_VL_SL}. Notably, the power-law indexes derived from different polarizations, namely $\gamma_{\ST}$, $\gamma_{\VL}$, and $\gamma_{\SL}$, exhibit broad consistency.
We also observe that the disfavored longitudinal modes remain slightly unfavorable when considered in combination with the TT mode. Specifically, we find $\mathrm{BF}^{\mathrm{TT+VL}}_{\mathrm{TT}} = 0.49$ and $\mathrm{BF}^{\mathrm{TT+SL}}_{\mathrm{TT}} = 0.27$.

%%%%%%%%%%%%%%%%%%%%%%%%%%%%%%%%%%%%%%%%%%%%%%%%%%%%%%%%%%%%%%%%%%%%%%%%%%%%%%%%

\begin{figure}[tbp!]
    \centering
    \includegraphics[width=\linewidth]{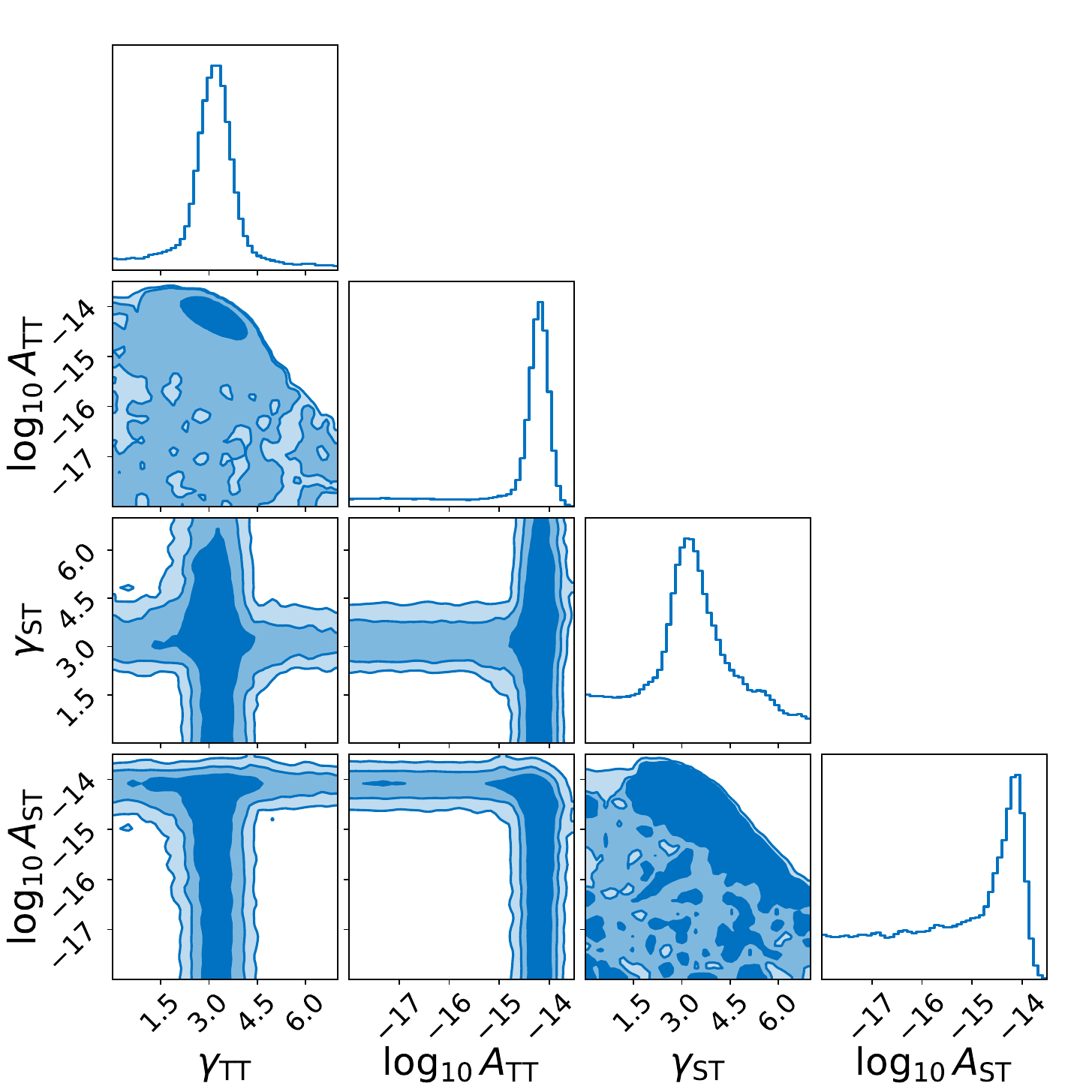}
    \caption{\label{post_TTST} Same as~\Fig{post_ST_VL_SL} but for the TT+ST model.}
\end{figure}

We also consider a TT+ST model that simultaneously incorporates both the TT and ST correlations. The Bayes factor between the TT+ST model and the TT model is $0.943\pm 0.005$, indicating that there is no significant evidence supporting or refuting the ST correlations in addition to the HD correlations. The contour plot and the posterior distributions of the TT and ST components in the TT+ST model are depicted in \Fig{post_TTST}. The presence of the peak around $-14.5$ for the amplitude $\log_{10} A_{\ST}$ confirms that the NANOGrav data do not rule out the possibility of an ST signal in the data.

\begin{figure}[tbp!]
	\centering
	\includegraphics[width=\linewidth]{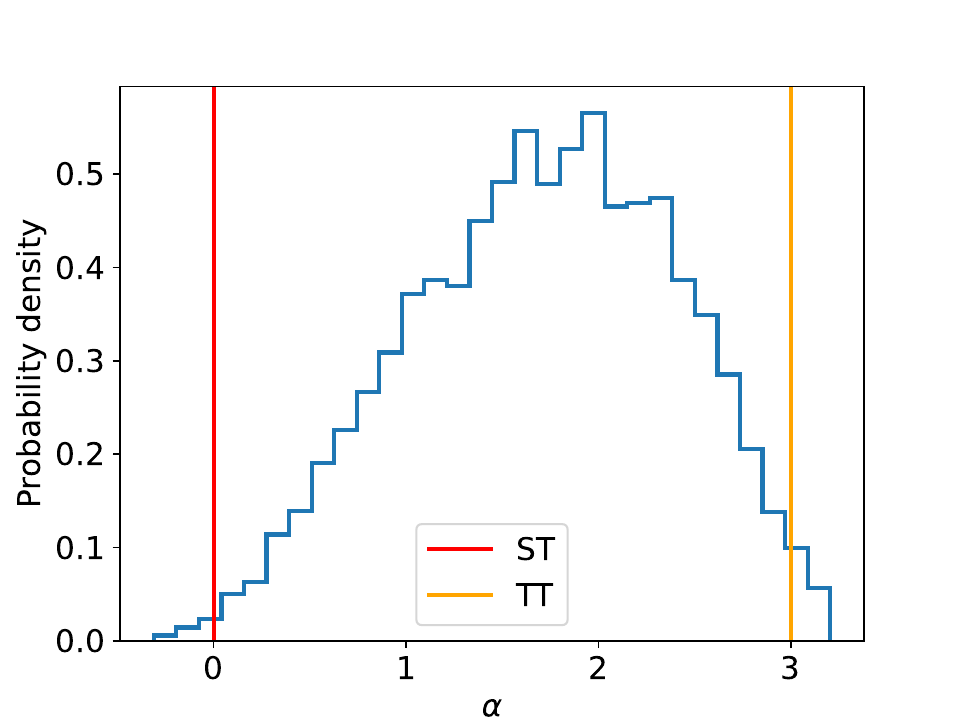}
	\caption{\label{post_GT} Bayesian posteriors for the $\alpha$ parameter in the GT model with a parameterized ORF as defined by~\Eq{alpha}. The two vertical dashed lines correspond to the ST ($\alpha=0$) and TT ($\alpha=3$) ORFs, respectively.}
\end{figure}

To gain further insights into the optimal correlations that best describe the data, we also fit the NANOGrav 15-year data set with a parameterized ORF (the GT model) as defined in \Eq{alpha}. The posterior distribution for the free parameter $\alpha$ is displayed in \Fig{post_GT}. Our analysis yields a value of $\alpha=1.74^{+1.18}_{-1.41}$, thus excluding both the TT and ST models at the $90\%$ confidence level. However, it is worth noting that the TT and ST models remain consistent with the NANOGrav 15-year data set at the $3\sigma$ confidence level. Furthermore, the Bayes factor between the GTb model, which is the GT model by fixing $\alpha$ to the best-fit value of $1.74$, and the TT model is $3.9\pm 0.3$, confirming that NANOGrav can be better described by the correlations with $\alpha\simeq 1.74$ than by the HD correlations with $\alpha=3$.
Furthermore, we calculate the Bayes factor comparing the TT+ST model and the GTb model as $\mathrm{BF}^{\mathrm{TT+ST}}_{\mathrm{GTb}} = 0.24$, indicating a preference for the GTb model over the TT+ST model.
% Even in the presence of a monopole or dipole signal, the GTb model consistently outperforms other polarization models as can be seen from \Table{bayes2}.}

More comprehensive model comparisons can be obtained when considering the presence of a monopole or dipole signal, as shown in both \Table{bayes} and \Table{bayes2}. Particularly, from \Table{bayes2}, it is evident that even when accounting for these additional signals, the GTb model consistently outperforms other polarization models.

% Even in the presence of a monopole or dipole signal, the GTb model consistently outperforms other polarization models as can be seen from \Table{bayes2}.}
% For example, the Bayes factor comparing the TT+ST model and the TT+monopole model is $\mathrm{BF}^{\mathrm{TT+ST}}_{\mathrm{TT+mono}} = 1.72$, indicating a slight preference for the TT+ST model over the TT+monopole model.
% Similarly, the Bayes factor comparing the GTb+monopole model and the TT+monopole model is $\mathrm{BF}^{\mathrm{GTb+mono}}_{\mathrm{TT+mono}} = 4.56$, suggesting that the GTb+monopole model provides a better fit to the data than the TT+monopole model.
% Additionally, the Bayes factor comparing the GTb+monopole+dipole model and the TT+monopole+dipole model is $\mathrm{BF}^{\mathrm{GTb+mono+dipole}}_{\mathrm{TT+mono+dipole}} = 1.02$, indicating that the GT+monopole+dipole model offers a better fit to the data than the TT+monopole+dipole model.

%%%%%%%%%%%%%%%%%%%%%%%%%%%%%%%%%%%%%%%%%%%%%%%%%%%%%%%%%%%%%%%%%%%%%%%%%%%%%%%%

In summary, our analysis indicates that the NANOGrav 15-year data set can be effectively described by either the TT correlations or the ST correlations, with no compelling evidence strongly favoring one over the other.
The best-fitting model, GT with $\alpha\simeq1.74$, might result from the potential impact of model misspecification, either in pulsar noise terms or the choice of model correlations.
Nevertheless, the current PTA data cannot provide a definitive verdict on the spatial correlations within the stochastic signal, thereby posing a challenge for the unequivocal detection of the GWB predicted by general relativity through PTAs. 
Future work may need to consider the potential impact of cosmic variance in this quest~\cite{Allen:2022dzg,Allen:2022ksj,Bernardo:2022xzl,Bernardo:2023pwt,Bernardo:2023bqx,Bernardo:2023zna}.
We anticipate that future PTA data with a longer observation timespan and a larger number of pulsars will provide the necessary insights to pinpoint the origin of the observed stochastic signal.

It is noteworthy that the official NANOGrav collaboration independently conducted a study on transverse modes in their 15-year data set, with their findings appearing on the arXiv shortly after ours~\cite{NANOGrav:2023ygs}. Their analysis revealed strong Bayes factors supporting a correlated signal. Notably, the data did not exhibit a strong preference for either correlation signature, with Bayes factors around $\sim 2$ when comparing TT to ST correlations and $\sim 1$ for TT+ST correlations against TT correlations alone. These outcomes align closely with our findings, indicating robust consistency. However, certain differences between our analyses emerge. The official NANOGrav collaboration performed dropout analysis tests, which we did not undertake. They identified J0030+0451 and J0613$-$0200 as primarily responsible for the ST significance. Upon excluding these pulsars, they observed a significant increase in the Bayes factor for HD, accompanied by a notable reduction in the Bayes factor for ST. In contrast, our study extends beyond theirs by including searches for longitudinal polarization modes (VL in Eq.~\eqref{gammaVL} and SL in Eq.~\eqref{gammaSL}), aspects not explored in their analysis. Furthermore, we delved into a GT polarization mode characterized by the $\alpha$ parameter (see Eq.~\eqref{alpha}), an avenue untouched by the official NANOGrav collaboration's investigation. Our findings indicate that the data can be best described by the GT correlations with $\alpha\simeq 1.74$.

\textbf{Note Added.} A similar study by the NANOGrav collaboration~\cite{NANOGrav:2023ygs}, which explores transverse polarization modes in their 15-year data set, was posted on arXiv one day after our manuscript. While the results regarding the ST mode from Ref.~\cite{NANOGrav:2023ygs} are largely consistent with our findings, it is worth highlighting that our study investigates the SL, VL, and GT models that were not examined in Ref.~\cite{NANOGrav:2023ygs}.

%%%%%%%%%%%%%%%%%%%%%%%%%%%%%%%%%%%%%%%%%%%%%%%%%%%%%%%%%%%%%%%%%%%%%%%%%%%%%%%%
%%%%%%%%%%%%%%%%%%%%%%%%%%%%%% acknowledgments %%%%%%%%%%%%%%%%%%%%%%%%%%%%%%%%%
%%%%%%%%%%%%%%%%%%%%%%%%%%%%%%%%%%%%%%%%%%%%%%%%%%%%%%%%%%%%%%%%%%%%%%%%%%%%%%%%
\emph{Acknowledgments}
% We would like to thank the anonymous referee for the useful suggestions and comments. 
We acknowledge the use of the HPC Cluster of ITP-CAS. QGH is supported by grants from NSFC (Grant No.~12250010, 11975019, 11991052, 12047503), Key Research Program of Frontier Sciences, CAS, Grant No.~ZDBS-LY-7009, the Key Research Program of the Chinese Academy of Sciences (Grant No.~XDPB15). 
ZCC is supported by the National Natural Science Foundation of China (Grant No.~12247176 and No.~12247112) and the China Postdoctoral Science Foundation Fellowship No.~2022M710429.
	
%%%%%%%%%%%%%%%%%%%%%%%%%%%%%%%%%%%%%%%%%%%%%%%%%%%%%%%%%%%%%%%%%%%%%%%%%%%%%%%%
%%%%%%%%%%%%%%%%%%%%%%%%%%%%%%%%%%%%% references %%%%%%%%%%%%%%%%%%%%%%%%%%%%%%%%%%%%
%%%%%%%%%%%%%%%%%%%%%%%%%%%%%%%%%%%%%%%%%%%%%%%%%%%%%%%%%%%%%%%%%%%%%%%%%%%%%%%%
%\bibliographystyle{apj}
\bibliography{./ref}

%%%%%%%%%%%%%%%%%%%%%%%%%%%%%%%%%%%%%%%%%%%%%%%%%%%%%%%%%%%%%%%%%%%%%%%%%%%%%%%%
\end{document}